\documentstyle[12pt]{article}

\newcommand{\setZ}{\ifmmode{{\bf Z}}
    \else{{\bf Z}}\fi} 
\newcommand{\setC}{\ifmmode{C\hskip-6pt\vrule height6.7pt\hskip6.7pt}
    \else{\hbox{$C\hskip-6pt\vrule height6.7pt\hskip6.7pt$}}\fi}
\newcommand{\BE}{\begin{equation}}
\newcommand{\EE}{\end{equation}}
\newcommand{\BD}{\begin{displaymath}}
\newcommand{\ED}{\end{displaymath}}
\newcommand{\BEA}{\begin{eqnarray}}
\newcommand{\EEA}{\end{eqnarray}}
\newcommand{\BEAS}{\begin{eqnarray*}}
\newcommand{\EEAS}{\end{eqnarray*}}
\newcommand{\alA}{\ifmmode {\cal A}\else $\cal A$\fi}
\newcommand{\alB}{\ifmmode {\cal B}\else$\cal B$\fi}
\newcommand{\ndots}{
   \unitlength0.8ex
   \begin{picture}(2,2)
   \put(0.3,0.3){.}
   \put(1,1){.}
   \put(1.7,1.7){.}
   \end{picture} }
\newcommand{\tr}{{\rm tr\,}}
\newcommand{\id}{{\rm id}}
\newcommand{\mod}{\,{\rm mod\,}}
\newtheorem{theorem}{Theorem}
\newcommand{\tprod}[3]{#1_{#2}\otimes\ldots\otimes #1_{#3}}
\newcommand{\ttprod}[3]{{#1}^{#2\ldots #3}e_{#2}\otimes\ldots\otimes e_{#3}}
\newcommand{\refeq}[2]{\stackrel{(\ref{#2})}{#1}}

\begin{document}
\begin{center}
{\Large Flip-Moves and Graded Associative Algebras}\\[1cm]
{\large Claus Nowak}\\[5mm]
Fakult\"at f\"ur Physik der Universit\"at Freiburg\\
Hermann-Herder-Str. 3, 79104 Freiburg i.Br. / FRG
\end{center}

\begin{abstract}
The relation between discrete topological field theories on triangulations
of two-dimensional manifolds and associative algebras was worked out
recently. The starting point for this development was the graphical
interpretation of the
associativity as flip of triangles. We show that there is
a more general relation between flip-moves with two $n$-gons
and $Z_{n-2}$-graded associative algebras. A detailed examination shows
that flip-invariant models on a lattice of $n$-gons can be constructed
{}from $Z_{2}$- or $Z_{1}$-graded algebras, reducing in the second case
to triangulations of the two-dimensional manifolds.
Related problems occure naturally in three-dimensional topological lattice
theories.
\end{abstract}

\hfill\begin{minipage}{5cm}
\begin{center}
University of Freiburg\\
July 1994\\
THEP 94/6
\end{center}
\end{minipage}
\vspace{2cm}

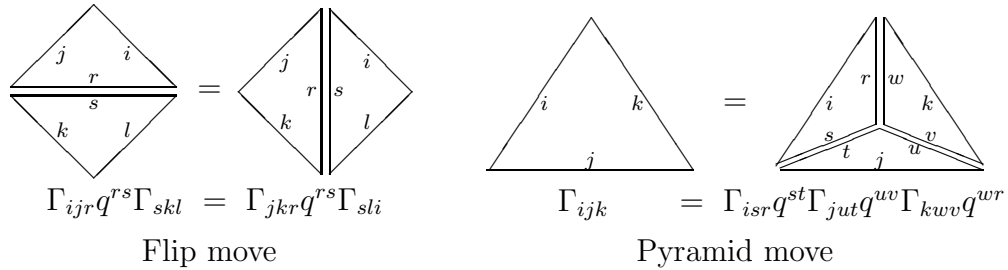
\begin{figure}[hbt]
\null\hfill
\unitlength=0.55mm
\begin{picture}(240.00,60.00)(0,-5)
\put(5.00,40.00){\line(1,1){20.00}}
\put(25.00,60.00){\line(1,-1){20.00}}
\put(45.00,40.00){\line(-1,0){40.00}}
\put(5.00,38.00){\line(1,0){40.00}}
\put(45.00,38.00){\line(-1,-1){20.00}}
\put(25.00,18.00){\line(-1,1){20.00}}
\put(60.00,39.00){\line(1,1){20.00}}
\put(80.00,59.00){\line(0,-1){40.00}}
\put(80.00,19.00){\line(-1,1){20.00}}
\put(82.00,59.00){\line(0,-1){40.00}}
\put(82.00,19.00){\line(1,1){20.00}}
\put(102.00,39.00){\line(-1,1){20.00}}
\put(53.00,39.00){\makebox(0,0)[cc]{$=$}}
\put(120.00,20.00){\line(1,0){50.00}}
\put(170.00,20.00){\line(-2,3){24.67}}
\put(145.33,57.00){\line(-2,-3){24.67}}
\put(190.00,21.00){\line(2,3){24.00}}
\put(214.00,57.00){\line(0,-1){26.00}}
\put(214.00,31.00){\line(-5,-2){24.00}}
\put(191.00,20.00){\line(1,0){49.00}}
\put(240.00,20.00){\line(-5,2){25.00}}
\put(215.00,30.00){\line(-5,-2){24.00}}
\put(216.00,57.00){\line(0,-1){26.00}}
\put(216.00,31.00){\line(5,-2){24.00}}
\put(240.00,21.33){\line(-2,3){23.67}}
\put(180.00,36.00){\makebox(0,0)[cc]{$=$}}
\scriptsize
\put(16.00,50.00){\makebox(0,0)[lt]{$j$}}
\put(34.00,50.00){\makebox(0,0)[rt]{$i$}}
\put(25.00,41.00){\makebox(0,0)[cb]{$r$}}
\put(25.00,37.00){\makebox(0,0)[ct]{$s$}}
\put(16.00,28.00){\makebox(0,0)[lb]{$k$}}
\put(34.00,28.00){\makebox(0,0)[rb]{$l$}}
\put(70.00,48.00){\makebox(0,0)[lt]{$j$}}
\put(70.00,30.00){\makebox(0,0)[lb]{$k$}}
\put(92.00,30.00){\makebox(0,0)[rb]{$l$}}
\put(92.00,48.00){\makebox(0,0)[rt]{$i$}}
\put(79.00,39.00){\makebox(0,0)[rc]{$r$}}
\put(83.00,39.00){\makebox(0,0)[lc]{$s$}}
\put(133.00,38.00){\makebox(0,0)[lt]{$i$}}
\put(158.00,38.00){\makebox(0,0)[rt]{$k$}}
\put(145.00,21.00){\makebox(0,0)[cb]{$j$}}
\put(202.00,38.00){\makebox(0,0)[lt]{$i$}}
\put(228.00,38.00){\makebox(0,0)[rt]{$k$}}
\put(215.00,21.00){\makebox(0,0)[cb]{$j$}}
\put(204.00,27.00){\makebox(0,0)[rb]{$s$}}
\put(213.00,42.00){\makebox(0,0)[rc]{$r$}}
\put(217.00,42.00){\makebox(0,0)[lc]{$w$}}
\put(206.00,26.00){\makebox(0,0)[lt]{$t$}}
\put(225.00,26.00){\makebox(0,0)[rt]{$u$}}
\put(226.00,27.00){\makebox(0,0)[lb]{$v$}}
\normalsize
\put(53.00,12.00){\makebox(0,0)[cc]{
$\Gamma_{ijr}q^{rs}\Gamma_{skl}~=~\Gamma_{jkr}q^{rs}\Gamma_{sli}$}}
\put(180.00,12.00){\makebox(0,0)[cc]{
$~~~~~~~~\Gamma_{ijk}~~~~~~=~\Gamma_{isr}q^{st}\Gamma_{jut}q^{uv}
\Gamma_{kwv}q^{wr}$}}
\put(53.00,0.00){\makebox(0,0)[cc]{Flip move}}
\put(180.00,0.00){\makebox(0,0)[cc]{Pyramid move}}
\end{picture} \hfill\null
\caption{\label{bothflips}Moves for $n=3$}
\end{figure}

Various aspects of topological lattice theories had been considered
in the last years. First models had been constructed as discrete analogies
of continuous topological field theories. The invariance of the continuous
theorie under the diffeomorphism group was discretised to the invariance
under flip moves of the lattice \cite{JonWhe}, see fig. \ref{bothflips}.
The field variables were
located on the vertices of the triangulation. Another type of models grow
out of matrix models of two-dimensional quantum gravity \cite{Filk},
where one wants to couple a topological action to the model to control the
topology-dependence of the series-expansion. These models have the field
variables on the edges of the triangles and could be classified by
associative algebras \cite{Bachas,BFN}. The approach to topological
lattice theories from the matrix models poses the problem to handle `
topological' actions coupled to models which not only contain a cubic but
higher polynoms in the potential. This was solved in \cite{BFN} for
monoms of degree 4, leading to quadrangulations of two-dimensional
manifolds, and for arbitrary polynoms containing a cubic term, this leads
to lattices build out of triangles and higher polygons.

This paper is a part of \cite{Nowak} and treats the remaining models for
monomials of arbitrary degree,
i.e. for manifolds covered by $n$-gons. This is of interest not only
in the two-dimensional case, in a special case of three-dimensional
topological lattice-theories \cite{CFS} one has to deal with polygonals
and multivalent hinges. For this the subdivision invariance of the weights
must be assumed, a condition which is in our work a consequence of a rather
natural condition to the weights.

We construct from the given data, the sets of weights
$\Gamma_{i_1\ldots i_n}$ of the $n$-gons and the weights $q^{ij}$ of the
edges, an associative graded algebra, which allows for the classification
and the computation of the partition function. We recover the topological
model on triangulations and the models on chequered graphs already found
in \cite{BFN} and show, that this is a complete classification, all
flip-invariant models on polygonizations belong to one of these two types.

First we have to introduce the model.
We consider a polygonization of a two-dimensional compact oriented manifold
by $n$-gons. On this polygonization we
establish a statistical model with variables $i,j,\ldots=1,\ldots,N$
on the edges of the $n$-gons, weights $\Gamma_{i_1\ldots i_n}\in\setC$
on the $n$-gons and $q^{ij}\in\setC$ on the edges. The weights $\Gamma$
have to be cyclic, the weights $q$ have to be symmetric.

We assume that the matrix $(q^{ij})$ is regular and the inverse matrix
$(q_{ij})$ exists (this condition can always be achieved by a simple
transformation and a reduction of the range of the indices, see \cite{BFN}).
The partition function is the sum of the product of all weights over
all indices.

For $n=3$ the model is called topological if the weights are invariant
under the moves in fig. \ref{bothflips}, these moves are transitive
on the set of all two-dimensional simplicial complexes, this was already
shown by Alexander in 1930 \cite{Alexander}, see also \cite{MGross} for a
discussion.

In this case one defines an
algebra $\cal A$ which is a vector space with basis $\{e_1,\ldots,e_N\}$
and multiplication $e_i\cdot e_j=\lambda_{ij}^k e_k$, where the
structure constants are formed by $\lambda_{ij}^k=\Gamma_{ijr}q^{rk}$
(Here and in the following , the sum convention is assumed). The elements
of the matrix $(q_{ij})$, the invers matrix of $(q^{ij})$, form the
coefficients of a symmetric bilinear form $q$ on $\cal A$ with
$q_{ij}=q(e_i,e_j)$. Due to the cyclicity of the $\Gamma_{ijk}$ this
bilinear form is invariant under multiplications in $\cal A$:
\BE  \label{3invariance}
    q(a\cdot b,c)~=~q(a,b\cdot c)~,~~~~~\forall~a,b,c\in\cal A
\EE
An algebra together with a metric which fulfils (\ref{3invariance}) is
called {\it metrised}, see e.g. \cite{Bordemann}.

As shown in several publications \cite{Bachas,BFN}, the conditions
imposed by the flip and the pyramid move make the algebra
associative and semisimple. The flip condition
is the cause for associativity, the pyramid condition
was thought to be the origin of the semisimplicity, but in \cite{BFN} it
was shown that the ``non-semisimple parts'' of the algebra (which consists
not only of the radical, but also of some Levi subalgebra) give no
contributions to the partition function of the statistical models considered
here, and can therefore be ignored. What remains is a semisimple algebra.
Imposing the pyramid flip is therefore not necessary for the classification
of topological models.

The relation between flip moves and associative algebras was (see
\cite{BFN}) extended to the case of flips of two 4-gons,
leading to $\setZ_2$-graded associative algebras. There occured the new
quality, that some of the models vanish on graphs which can not be
chequered.

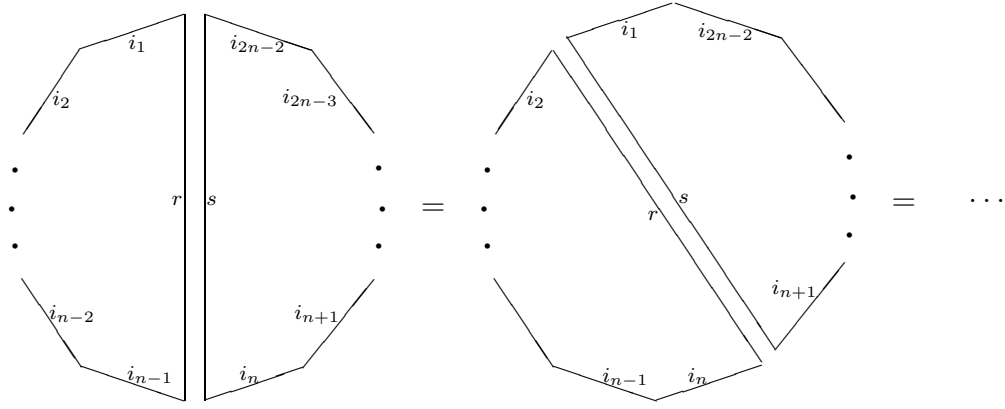
\begin{figure}[hbt]
\null\hfill
\unitlength=2.7mm
\begin{picture}(50.50,20.58)
\put(3.00,7.00){\line(2,-3){2.92}}
\put(5.92,2.67){\line(3,-1){5.08}}
\put(11.00,1.00){\line(0,1){19.00}}
\put(11.00,20.00){\line(-3,-1){5.17}}
\put(5.83,18.25){\line(-2,-3){2.75}}
\put(12.00,20.00){\line(0,-1){19.00}}
\put(12.00,1.00){\line(3,1){4.83}}
\put(16.83,2.58){\line(4,5){3.50}}
\put(12.00,20.00){\line(3,-1){5.25}}
\put(17.25,18.25){\line(3,-4){3.08}}
\put(2.50,10.42){\circle*{0.24}}
\put(2.67,12.33){\circle*{0.24}}
\put(2.67,8.58){\circle*{0.24}}
\put(20.75,10.42){\circle*{0.24}}
\put(20.58,12.42){\circle*{0.24}}
\put(20.58,8.58){\circle*{0.24}}
\put(26.25,7.00){\line(2,-3){2.92}}
\put(29.17,2.67){\line(3,-1){5.08}}
\put(35.00,20.58){\line(-3,-1){5.17}}
\put(29.08,18.25){\line(-2,-3){2.75}}
\put(35.17,20.58){\line(3,-1){5.25}}
\put(40.42,18.83){\line(3,-4){3.08}}
\put(25.75,10.42){\circle*{0.24}}
\put(25.92,12.33){\circle*{0.24}}
\put(25.92,8.58){\circle*{0.24}}
\put(43.92,11.00){\circle*{0.24}}
\put(43.75,13.00){\circle*{0.24}}
\put(43.75,9.16){\circle*{0.24}}
\put(29.17,18.25){\line(2,-3){10.25}}
\put(34.25,1.00){\line(3,1){5.17}}
\put(29.83,18.92){\line(2,-3){10.25}}
\put(40.08,3.50){\line(4,5){3.42}}
\put(23.25,10.42){\makebox(0,0)[cc]{$=$}}
\put(46.42,10.67){\makebox(0,0)[cc]{$=$}}
\put(50.50,10.83){\makebox(0,0)[cc]{$\ldots$}}
\scriptsize
\put(8.25,19.00){\makebox(0,0)[lt]{$i_1$}}
\put(4.50,16.25){\makebox(0,0)[lt]{$i_2$}}
\put(4.33,5.00){\makebox(0,0)[lb]{$i_{n-2}$}}
\put(8.17,2.00){\makebox(0,0)[lb]{$i_{n-1}$}}
\put(10.92,10.83){\makebox(0,0)[rc]{$r$}}
\put(12.08,10.83){\makebox(0,0)[lc]{$s$}}
\put(14.75,2.00){\makebox(0,0)[rb]{$i_n$}}
\put(18.67,5.00){\makebox(0,0)[rb]{$i_{n+1}$}}
\put(16.00,19.00){\makebox(0,0)[rt]{$i_{2n-2}$}}
\put(18.58,16.33){\makebox(0,0)[rt]{$i_{2n-3}$}}
\put(27.83,16.25){\makebox(0,0)[lt]{$i_{2}$}}
\put(31.58,1.92){\makebox(0,0)[lb]{$i_{n-1}$}}
\put(36.83,1.92){\makebox(0,0)[rb]{$i_n$}}
\put(42.17,6.25){\makebox(0,0)[rb]{$i_{n+1}$}}
\put(39.05,19.58){\makebox(0,0)[rt]{$i_{2n-2}$}}
\put(32.50,19.67){\makebox(0,0)[lt]{$i_1$}}
\put(34.33,10.33){\makebox(0,0)[rt]{$r$}}
\put(35.33,10.75){\makebox(0,0)[lb]{$s$}}
\normalsize
\end{picture}\hfill\null
\caption{\label{nflips}Flip moves for two $n$-gons}
\end{figure}

We now generalize the work in \cite{BFN} to arbitrary $n$-gons.
First we generalize the flip move in fig. \ref{bothflips} for two $n$-gons
as shown in fig. \ref{nflips}. Imposing a condition similar to the pyramid
move will not be necessary for the classification of the topological models.

The weights invariant under the moves in fig. \ref{nflips} fulfil the
relations
\BD
   \Gamma_{i_1\ldots i_{n-1}r}q^{rs}\Gamma_{si_n\ldots i_{2n-2}} ~=~
   \Gamma_{i_2\ldots i_n r}q^{rs}\Gamma_{s i_{n+1}\ldots i_{2n-2}i_1}~=~
   \ldots
\ED
Trivial examples of weights invariant under these flips are constructed
out of models on
triangulations, the weight of the $n$-gon is defined by the fusion of
the weights of $n-2$ triangles. A nontrivial example is the four-vertex
model which was discussed in \cite{Filk}. We will see that all models
are analogous to one of these examples.

As in the case $n=3$ we define a $N$-dimensional complex vector space
$\cal A$ with basis $\{e_1,\ldots,e_N\}$ and a metric $q$ on ${\cal A}$ by
$q(e_i,e_j)=q_{ij}$. We define a $(n-1)$-linear map
$\Gamma:{\cal A}\times\ldots\times{\cal A}\mapsto{\cal A}$ by
\BD
    \Gamma(e_{i_1},\ldots,e_{i_{n-1}})~
         :=~\Gamma_{i_1\ldots i_{n-1}r}q^{rs}e_s~~.
\ED

Again the metric is invariant with respect to the map $\Gamma$:
\BE \label{geninv}
\begin{array}{rcl}
   q(\Gamma(e_{i_1},\ldots,e_{i_{n-1}}),e_{i_n}) &=&
   \Gamma_{i_1\ldots i_n}~=~\Gamma_{i_2\ldots i_n i_1}\\
    &=&q(\Gamma(e_{i_2},\ldots,e_{i_n}),e_{i_1})\\
    &=& q(e_{i_1},\Gamma(e_{i_2},\ldots,e_{i_n}))~~.
\end{array}
\EE

The flip condition in fig. \ref{nflips} imposes the following conditions on
the map $\Gamma$:
\BEA   \nonumber
  \Gamma(\Gamma(a_1,\ldots,a_{n-1}),a_n,\ldots,a_{2n-3}) &=&
         \Gamma(a_1,\Gamma(a_2,\ldots,a_n),a_{n+1},\ldots,a_{2n-3})\\
   =~\ldots&=&\Gamma(a_1,\ldots,a_{n-2},\Gamma(a_{n-1},\ldots,a_{2n-3}))
    \label{genass}
\EEA
which are equivalent to
\BE
  \Gamma\circ(\id^{r}\otimes\Gamma\otimes \id^{n-2-r})~=~
  \Gamma\circ(\id^{s}\otimes\Gamma\otimes \id^{n-2-s})
\EE
for all $r,s=0,\ldots,n-2$.
This is a generalization of the associativity condition of associative
algebras.
An easy but time-consuming induction shows, that this general associativity
holds for more than two $\Gamma$:
\BE \label{allgvertausch}
  \begin{array}{ccc} \multicolumn{2}{c}{
  \Gamma\circ(\id^{r_1}\otimes\Gamma\otimes \id^{n-2-r_1})\circ\ldots\circ
   (\id^{r_k}\otimes\Gamma\otimes\id^{k(n-2)-r_k})~=~} & ~~~~~~~~~~~~\\
  {}~~~~~~~ & \multicolumn{2}{c}{ =~
  \Gamma\circ(\id^{s_1}\otimes\Gamma\otimes \id^{n-2-s_1})\circ\ldots\circ
   (\id^{s_k}\otimes\Gamma\otimes\id^{k(n-2)-s_k}) }
  \end{array}
\EE
for all admissible $r_i,s_i$.

For practical reasons we rename the vector space $\cal A$ by ${\cal A}_1$
and the metric $q$ by $q_1$.

Then we can prove the following main theorem:
\begin{theorem}  \label{allgstruktur}
Let ${\cal A}_1$ be a $N$-dimensional complex vector space.
Let $\Gamma : {\cal A}_1^{\times n-1}\to {\cal A}_1$ be a
$\setC$-multilinear map and $q_1:{\cal A}_1\times{\cal A}_1\to\setC$
a symmetric, non-degenerate metric, which satisfy the invariance
condition (\ref{geninv}) and the general associativity condition
(\ref{genass}).

Then there exists a $\setZ_{n-2}$-graded, associative, metrised
algebra $(\alA={\cal A}_0\oplus{\cal A}_1\oplus\ldots
\oplus{\cal A}_{n-3},q)$,
where $q$ is a non degenerate, symmetric bilinear form on ${\cal A}$ with
$q|_{{\cal A}_1\times{\cal A}_1}=q_1$ and
$q|_{{\cal A}_{i}\times{\cal A}_{j}}=0$ for $i+j\not\equiv 2\;\mod\;(n-2)$.

The map $\Gamma$ and the algebra multiplication are related by
\BE  \label{factors}
   \Gamma(a_1,\ldots,a_{n-1})~=~a_1\cdot\ldots\cdot a_{n-1}~~~~
           \forall~a_1,\ldots,a_{n-1}\in{\cal A}_1~~.
\EE
\end{theorem}

{\sc Remarks:}
An algebra ${\cal A}=\oplus_{i=0}^{m-1} {\cal A}_i$ is $\setZ_m$-graded,
if the multiplication fulfils ${\cal A}_i\times{\cal A}_j\to
{\cal A}_{i+j \mod m}$. Hence $a_1\cdot a_2\in\alA_2$,
$a_1\cdot a_2\cdot a_3\in\alA_3\ldots$, and finaly
$a_1\cdot\ldots\cdot a_n-1\in\alA_1$. We remark, that the algebra is not
super-graded, as it is assumed automatically by Lie-algebras.

Let $\{e_1,\ldots,e_{|{\cal A}|}\}$ be a ordered basis
of ${\cal A}$ with respect to the grading.
Let $\lambda_{ij}^k$ be the structure constants with respect
to this basis. Then we get with (\ref{factors}):
\BEA
     \Gamma_{i_1\ldots i_n} &=& q_1(\Gamma(e_{i_1},\ldots,e_{i_{n-1}})
         ,e_{i_n}) ~=~
      q(e_{i_1}\cdot\ldots\cdot e_{i_{n-1}},e_{i_n})\nonumber \\
      &=& q_{ri_n}\lambda_{i_1i_2}^{r_1}\lambda_{r_1i_3}^{r_2}\ldots
             \lambda_{r_{n-3}i_{n-1}}^r \nonumber \\
      &=& \lambda_{i_1 i_2 r_1}q^{r_1s_1}\lambda_{s_1i_3r_2}q^{r_2s_2}
          \ldots q^{r_{n-2}s_{n-2}}\lambda_{s_{n-2}i_{n-1} i_{n}}
      \label{zerleg}
\EEA

The inner indices are summed over $1,\ldots,\dim {\cal A}$, but due to
the grading of the algebra every summation is restricted to the indices
belonging to one part of the grading.

\begin{figure}[hbt]
\null\hfill
\unitlength=3.00mm
\begin{picture}(28.33,14.34)(0,3)
\put(3.08,9.50){\line(2,-5){1.83}}
\put(4.92,4.83){\line(6,1){5.42}}
\put(10.33,5.75){\line(2,3){2.83}}
\put(13.17,10.00){\line(-3,5){2.33}}
\put(10.83,13.92){\line(-1,0){5.17}}
\put(5.67,13.92){\line(-3,-5){2.67}}
\put(17.25,8.89){\line(2,-5){1.83}}
\put(19.08,4.33){\line(6,1){5.42}}
\put(20.08,13.75){\line(-3,-5){2.67}}
\put(20.83,14.33){\line(2,-1){7.50}}
\put(26.08,14.33){\line(3,-5){2.25}}
\put(20.50,13.92){\line(2,-1){7.58}}
\put(17.25,8.92){\line(2,-1){7.33}}
\put(15.25,10.00){\makebox(0,0)[cc]{$\rightarrow$}}
\put(20.81,14.34){\line(1,0){5.29}}
\put(17.40,9.28){\line(2,-1){7.62}}
\put(20.08,13.76){\line(3,-5){4.99}}
\put(28.09,10.11){\line(-2,-3){2.80}}
\put(20.50,13.90){\line(3,-5){4.79}}
\end{picture}
\hfill\null
\caption{Splitted $n$-gon\label{polsplit}}
\end{figure}
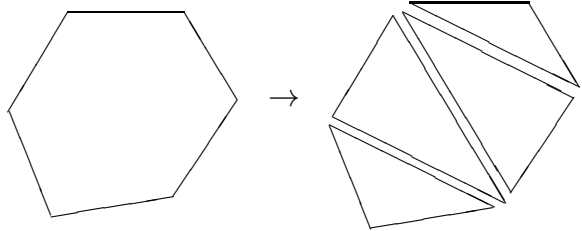

Thanks to the associativity we can replace the right hand side of
(\ref{zerleg}) by any evaluation of the associative product in
$q(e_{i_1}\cdot\ldots\cdot e_{i_{n-1}},e_{i_n})$. The graphical
interpretation is simple: we can replace the $n$-gon with weight
$\Gamma_{i_1\ldots i_n}$ by a triangulation with $n-2$ triangles
and weights $\lambda_{ijk}$ and sumation over all inner indices,
as in fig. \ref{polsplit}.
Due to the associativity of the algebra ${\cal A}$ is this model
flip invariant. We see: flip invariant models on $n$-gonisations of
a two-dimensional manifold are equivalent to flip invariant models
on triangulations with a greater range of indices and the
restriction, that certain indices only take values in the original
part. The value $n-2$ will appear often, therefore we define $p:=n-2$.

To prove the theorem we have to perform the following steps:
\begin{enumerate}
\item We define a non-associative algebra structure on the vectorspace
  $M=\oplus_{k=1}^{p}
  {\cal A}_1^{\otimes k}$ by the multiplication
  \BE   \label{multdef}
    a \cdot b~=~
    \left\{\begin{array}{ll}
    a_1\otimes\ldots\otimes a_k\otimes b_1\otimes\ldots\otimes b_l~=~
     a\otimes b& k+l\leq p\\
    \Gamma(a_1,\ldots,b_{p+1-k})\otimes b_{p+2-k}
    \otimes\ldots\otimes b_l &  k+l>p
    \end{array} \right.
  \EE
  for $a=\tprod a1k,~b=\tprod b1l$. This algebra is finite dimensional, but
  non-associative. The properties of $\Gamma$ allow the definition of an
  ideal $I$, such that $M/I$ is associative. This is not the usual way to
  construct an associative algebra, which would start with the infinite
  dimensional universal
  tensor algebra over $\alA_1$ and divide
  out an infinite dimensional ideal to get a finite dimensional associative
  algebra.
\item We define the subspace $I:=\oplus_{k=1}^{p}I_k$ of $M$
  with $I_1:=\{0\}$ and
  \BE  \label{idealdef}
    I_k~:=~\{a\in{\cal A}_1^{\otimes k}| a\cdot b=0~\forall~b\in
    {\cal A}_1^{p+1-k}\}
  \EE
  for $k=2,\ldots,p$ and show that $I$ is a two sided ideal of $M$.
\item We can therefore define the algebra
  \BE  \label{algebradef}
  {\cal A}~:=~M/I~=~\mathop{\oplus}\limits_{k=1}^{p}
         {\cal A}_1^{\otimes k}/I_k~=:~
              \mathop{\oplus}\limits_{k=1}^{p}{\cal A}_k
  \EE
  which will be shown to be associative and contains the original
  vector space ${\cal A}_1$.
  The relation (\ref{factors}) concides with the definiton of the
  multiplication.
\item We define a bilinear form $q$ on $M$ for $a=\tprod a1k,
  b=\tprod b1l\in M$ by
  \BE \label{metricdef}
    q(a,b):=\left\{
    \begin{array}{ll}
    0 & k+l\not\equiv 2\;\mod\; p \\
    q_1(a_1,\Gamma(a_2,\ldots,b_l))& k+l=n\\
    q_1(a_1,b_1) & k=l=1
    \end{array}  \right.
  \EE
  and show that the projection of $q$ on \alA, which we will denote
  also by $q$, is well-defined and symmetric.
\item We show that $q$ is non-degenerate on $\cal A$.
\item We show that $q$ is invariant on $\cal A$.
\end{enumerate}

Proofs and Remarks:

1) The multiplication (\ref{multdef}) is in general not associative,
consider e.g.
\BEAS
  a\cdot(\tprod b1{p}\cdot c) &=& a\otimes\Gamma(b_1,\ldots,b_p,c)\\
  (a\cdot\tprod b1{p})\cdot c &=& \Gamma(a,b_1,\ldots,b_p)\otimes c~~.
\EEAS
But we can show, that the multiplication is associative for factors
$a\in{\cal A}_1^{\otimes n_a},~b\in{\cal A}_1^{\otimes n_b},
{}~c\in{\cal A}_1^{\otimes n_c},\ldots$
with $n_a+n_b+n_c+\ldots\equiv 1 \mod p$, e.g.
\BEA
   (a\cdot b)\cdot c &=& a\cdot(b\cdot c)   \label{3prodass}\\
   ((a\cdot b)\cdot c)\cdot d &=& (a\cdot(b\cdot c))\cdot d
   \label{4prodass}
\EEA
For this, let $a=\tprod a1{n_a},~b=\tprod b1{n_b},~c=\tprod c1{n_c},\ldots~$.
Since $n_a+n_b+n_c+\ldots\equiv 1 \mod p$, all the products are of the
form
\BD
  \Gamma\circ(\id^{r_1}\otimes\Gamma\otimes\id^{p-r_1})\circ
  (\id^{r_2}\otimes\Gamma\otimes\id^{2p-r_2})\circ\ldots
  (a_1,\ldots,a_{n_a},b_1,\ldots)~~
\ED
and products of the same factors are equivalent by (\ref{allgvertausch}).

2) The condition $a\cdot b=0$ for all $b\in{\cal A}_1^{\otimes p+1-k}$ is
equivalent to $b\cdot a=0$ for all $b\in{\cal A}_1^{\otimes p+1-k}$ and
we can define alternatively
\BE  \label{altidealdef}
  I_k~=~\{a\in{\cal A}_1^{\otimes k}| b\cdot a=0~\forall~b\in
    {\cal A}_1^{\otimes p+1-k}\}
\EE
In order to see this, we use the invariance condition (\ref{geninv})
and the symmetry of $q_1$:
For all $b=b_1\otimes\ldots\otimes b_{p+1-k}\in{\cal A}_1^{\otimes
p+1-k}$, for all $c\in{\cal A}_1$ and for $a=a^{i_1\ldots i_k}
\tprod e{i_1}{i_k}\in I_k$ holds
\BEAS
  0 &=& a\cdot b~=~a^{i_1\ldots i_k}\Gamma(e_{i_1},\ldots,e_{i_k},b_1,
        \ldots,b_{p+1-k})\\
  \Leftrightarrow 0  &=& q_1(a^{i_1\ldots i_k}\Gamma(e_{i_1},
        \ldots,e_{i_k},b_1,\ldots,b_{p+1-k}),c)\\
  \Leftrightarrow 0  &\refeq{=}{geninv}&
        q_1(a^{i_1\ldots i_k}\Gamma(b_2,\ldots,b_{p+1-k}
        ,c,e_{i_1},\ldots,e_{i_k}),b_1)\\
  \Leftrightarrow 0  &=&a^{i_1\ldots i_k}\Gamma(b_2,\ldots,b_{p+1-k}
        ,c,e_{i_1},\ldots,e_{i_k})\\
   \Leftrightarrow 0  &=&(\tprod b1{p+1-k}\otimes c)\cdot a~~.
\EEAS
Since the elements of the form $\tprod b1{p+1-k}\otimes c$ span
${\cal A}_1^{\otimes p+1-k}$, the equivalence is proved.

The subspace $I$ of $M$ is an ideal of
$M$:  Let $a\in I_{n_a}$, then for all $b\in{\cal A}_1^{\otimes n_b},
{}~n_b=1,\ldots,p$
and all $c\in{\cal A}_1^{\otimes n_c}$ with $n_c$ such, that $n_a+n_b+n_c
\equiv 1\mod p$, we have
\BE  \label{rechtsinI}
   (a\cdot b)\cdot c~\refeq{=}{3prodass}~ a\cdot(b\cdot c)~=~0~~\Rightarrow
   a\cdot b\in I
\EE
since $b\cdot c\in{\cal A}_1^{\otimes p+1-k}$ and
\BD
   (b\cdot a)\cdot c~\refeq{=}{3prodass}~ b\cdot
        \underbrace{(a\cdot c)}_{\mbox{$\in I$ by (\ref{rechtsinI}})}~=~0
\ED
by (\ref{altidealdef}), hence also $b\cdot a\in I$ and $I$ is a two sided
ideal.

3) The algebra defined in (\ref{algebradef}) is associative.
To prove this, we have to show that for $a,b,c\in M$ holds
$(a\cdot b)\cdot c-a\cdot(b\cdot c)\in I$.
Again it is sufficient, to consider $a\in{\cal A}_1^{\otimes {n_a}},~
b\in{\cal A}_1^{\otimes {n_b}},~c\in{\cal A}_1^{\otimes {n_c}}$.
Let $d\in{\cal A}_1^{\otimes {n_d}}$ with $n_d$ such that
$n_a+n_b+n_c+n_d\equiv 1\,\mod p$. Then
\BEAS
   ((a\cdot b)\cdot c)\cdot d &\refeq{=}{4prodass}&
                 (a\cdot(b\cdot c))\cdot d\\
   \Rightarrow~~ && ((a\cdot b)\cdot c - a\cdot(b\cdot c))\cdot d~=~0
\EEAS
and since this holds for all $d$, the difference is in $I$ and the
algebra $\cal A$ is associative.

We define ${\cal A}_0:={\cal A}_{p}$. With the multiplication
${\cal A}_{i}\cdot{\cal A}_{j}\to{\cal A}_{i+j\;\mod\;p}$ becomes
$\alA={\cal A}_0\oplus\ldots\oplus
{\cal A}_{n-3}$ a $\setZ_{p}$-graded associative algebra.
The condition (\ref{factors}) is satisfied due to the definition
(\ref{multdef}):
Let $a_1,\ldots,a_{p+1}\in{\cal A}_1$. Then
$$
  a_1\cdot\ldots\cdot a_{p+1}~=~ \Gamma(a_1,\ldots,a_{p+1})~~.
$$

4) The map $q: M\times M\to \setC$ defined in (\ref{metricdef})
is well defined on ${\cal A}\times {\cal A}$.
To prove this we have to show that $q(a,b)$ is independent of the
choice of the representatives of $a$ and $b$, i.e. for all $c_a,c_b\in I$
holds: $q(a+c_a,b+c_b)=q(a,b)$, i.e. $q(a,c_b)=0=q(c_a,b)=q(c_a,c_b)$.
Due to the block structure of $q$ it is sufficient to consider for $a,~b,~
c_a$ and $c_b$ only homogeneous elements.
Let $a=\tprod a1k\in{\alA}_1^{\otimes k}$,
$c_b=\ttprod c{j_1}{j_l}\in I_l$, $k+l\equiv 2\;\mod\;p$. In the case
$k=l=1$ $c_b=0$ and $q(a,c_b)=0$, for $k+l=n$ we have
\BEAS
   q(a,c_b) &=& q(\tprod a1k,\ttprod c{j_1}{j_l})\\
      &=& q_1(a_1,c^{j_1\ldots j_l}\Gamma(a_2,\ldots,a_k,
          e_{j_1},\ldots,e_{j_k}))\\
      &=& q_1(a_1,\underbrace{(\tprod a2k)\cdot c_b}_{=0})~=~0~~.
\EEAS
analogly $q(c_a,b)=0$, $q(c_a,c_b)=0$ is then clear.

$q$ is symmetric: for $a=\tprod a1k$, $b=\tprod b1l$ is
\begin{itemize}
\item $k+l\not\equiv 2\;\mod\;p$ : $q(a,b)=0=q(b,a)$
\item $k=l=1$ : $q(a,b)=q_1(a,b)=q_1(b,a)=q(b,a)$
\item $k+l=n$ :
\BEAS
    q(a,b) &=& q(\tprod a1k,\tprod b1l) ~=~
       q_1(a_1,\Gamma(a_2,\ldots,a_k,b_1,\ldots,b_l))\\
       &\refeq{=}{geninv}&
           q_1(b_1,\Gamma(b_2,\ldots,b_l,a_1,\ldots,a_k))~=~q(b,a)\\
\EEAS
\end{itemize}
5) $q$ is non-degenerate: Due to the block structure of the metric it is
again sufficient to consider homogeneous elements
$a\in{\cal A}_1^{\otimes k}$ and
$c=c^{i_1\ldots i_l}\tprod{e}{i_1}{i_l}\in{\cal A}_1^{\otimes l}$,
$k+l\equiv 2\;\mod\;p$.
Let $q(a,c)=0$ for all $a=\tprod a1k\in{\alA}_1^{\otimes k}$:
\BEAS
  \Leftrightarrow && q(\tprod a1k,\ttprod c{j_1}{j_l})=0~~\forall~a_1,\ldots
        a_k\in{\cal A}_1\\
  \Leftrightarrow &&  q_1(a_1,c^{j_1\ldots j_l}
    \Gamma(a_2,\ldots,a_k,e_{j_1},\ldots,e_{j_l}))~=~0~~
       \forall~a_1,\ldots,a_k\in{\cal A}_1\\
  \Leftrightarrow && c^{j_1\ldots j_l}\Gamma(a_2,\ldots,a_k,e_{j_1},
       \ldots,e_{j_l})~=~(a_2\otimes\ldots\otimes a_k)\cdot c~=~0~~\\
  \Leftrightarrow &&c\in I_l~.
\EEAS

6) To prove the invariance of the metric we consider again
$a=\tprod a1{n_a},~b=\tprod b1{n_b},
c=\tprod c1{n_c}$. For $n_a+n_b+n_c\not\equiv 2\,\mod\,p$ we have
$$
   q(a,b\cdot c)~=~0~=~q(a\cdot b,c)~~.
$$
For $n_a+n_b+n_c=n$ we have
\BEA
    q(a\cdot b,c) &=& q(c,a\cdot b)~=~q_1(c_1,\Gamma(c_2,\ldots,
       c_{n_c},a_1,\ldots,a_{n_a},b_1,\ldots,b_{n_b})) \nonumber\\
          &=& q_1(a_1,\Gamma(a_2,\ldots,a_{n_a},b_1,\ldots,b_{n_b},
                  c_1,\ldots,c_{n_c}))~=~ q(a,b\cdot c)  \label{inv4n}
\EEA
For $n_a+n_b+n_c\equiv 2\,\mod\, p$ let $a'=\tprod a2k$, i.e.
$a=a_1\cdot a'$. Then
\BEAS
   q(a\cdot b,c)&=& q((a_1\cdot a')\cdot b,c)~=~q(a_1\cdot(a'\cdot b),c)\\
  &\refeq{=}{inv4n} & q(a_1,(a'\cdot b)\cdot c)~=~q(a_1,a'\cdot(b\cdot c))\\
  &\refeq{=}{inv4n} & q(a_1\cdot a',b\cdot c) ~=~ q(a,b\cdot c)
\EEAS

The metric $q$ has a block structure with respect to the decomposition
$\alA=\oplus_{k=0}^{n-3}{\cal A}_{k}$, due to
$q({\cal A}_i,{\cal A}_j)=0$ for $i+j\not\equiv 2\;\mod\;p$ we get
for $n>4$
\BE
q~=~(q_{ij})~=~\left(\begin{array}{cccccc}
          0&&\Box&&& \\
          &\Box&&&0& \\
          \Box &&&&&\\
          &&&&&\Box \\
          &0&&&\ndots&\\
          &&&\Box&&0
         \end{array}\right)
\EE
The $i$-th column and row, respectively, belong to the component
${\cal A}_{i-1}$ of \alA. We use the symbol $q$ for the metric and for the
matrix $(q_{ij})$ in a basis. We assume in the following, that we have
chosen a basis $\{e_i\}$ which respects the grading of \alA.
It is easy to see, that the inverse matrix $(q^{ij})$ has then same
structure; matrix elements $q^{ij}$ are only not equal zero, if
the basis elements $e_i$ and $e_j$ lie in components ${\cal A}_k$
and
${\cal A}_l$ with
$k+l\equiv 2\;\mod\;p$.~~$\Box$

We now use the methods elaborated in \cite{BFN} to calculate the
partition functions of the flip invariant models. We first review a
few facts about associative, metrised algebras (see \cite{BFN} for details).
\begin{itemize}
\item Let \alA\  be a complex, associative, metrised algebra. We decompose
   $\alA=\alB\oplus L\oplus R$, where \alB\  is the largest semisimple
   ideal of \alA, $L$ is a (non-unique) semisimple Levi-subalgebra and $R$
   is the radical of \alA. \alB\  and $L\oplus R$ are orthogonal with
   respect to $q$, i.e. $L\oplus R=\alB^\bot$.
\item $q^{ij}\not=0$ for $e_i\in L$ is only possible if $e_j\in R$
\item \alB\  itself is the direct sum of the simple ideals of \alA,
   $\alB=\oplus_i I_i$, where $I_i$ are the simple ideals of \alA and
   these are all orthogonal: $I_i \perp I_j$ for $i\not= j$.
\end{itemize}

We now check the relation of the decompositions $\alA=\alB\oplus L\oplus R$
and $\alA=\oplus_k \alA_k$. For this end we introduce the grading operator
$\theta$ on \alA\ by $\theta(a_k)=\omega^k a_k$ for $a_k\in\alA_k$, $\omega=
\exp(2\pi i/p)$. $\theta$ is an automorphism of \alA\ since
$\alA_k\times \alA_l\to\alA_{k+l\mod p}$.

Every $\theta$-invariant subalgebra $X$ of \alA\ allows for a decomposition
$X=\oplus_k X_k$ with $X_k\subset \alA_k$. \alB\ is a $\theta$-invariant
subalgebra, since the image of a semisimple ideal is a semisimple ideal,
therefore $\alB=\oplus_k \alB_k$. By the theorem 1 in \cite{Taft} there
exists a $\theta$-invariant Levi algebra $L$ which allows for a decomposition
$L=\oplus_k L_k$. The image of the radical $R$ is the radical, hence we have
also $R=\oplus_k R_k$.

We have therefore a decomposition of each $\alA_k=\alB_k\oplus L_k\oplus R_k$
and we can choose a basis of \alA\ respecting this decomposition.

Since \alB\ and $L\oplus R$ are orthogonal the partition function splits into
the partition function of a model with the semisimple algebra \alB\ and of
the algebra $L\oplus R$. The latter can be shown to be zero, the arguments
are the same as in \cite{BFN}, we will only give a sketch of the discussion.

\begin{figure}[hbt]
\null\hfill
\unitlength=2mm
\begin{picture}(19.53,25.08)(0,3)
\put(2.00,20.00){\line(0,-1){10.00}}
\put(9.50,2.50){\line(1,0){7.00}}
\put(2.00,20.00){\line(1,1){5.08}}
\put(2.00,20.00){\line(3,-1){12.00}}
\put(14.00,16.00){\line(-2,-1){12.00}}
\put(14.00,16.00){\line(-1,-3){4.50}}
\put(2.00,10.00){\line(1,-1){7.50}}
\put(14.00,16.00){\line(1,-2){2.92}}
\put(14.00,16.00){\line(-1,3){1.92}}
\put(2.25,15.25){\makebox(0,0)[lc]{$i_1$}}
\put(7.55,17.90){\makebox(0,0)[rt]{$r_1$}}
\put(7.55,13.00){\makebox(0,0)[rb]{$s_1$}}
\put(7.75,12.50){\makebox(0,0)[lt]{$r_2$}}
\put(6.00,6.42){\makebox(0,0)[lb]{$i_2$}}
\put(11.42,9.25){\makebox(0,0)[rb]{$s_2$}}
\put(12.00,9.17){\makebox(0,0)[lt]{$r_3$}}
\put(14.67,3.00){\makebox(0,0)[cb]{$i_3$}}
\put(7.00,18.70){\makebox(0,0)[lb]{$s_N$}}
\put(5.83,23.50){\makebox(0,0)[lt]{$i_N$}}
\put(10.08,22.08){\circle*{0.4}}
\put(11.83,23.08){\circle*{0.4}}
\put(13.83,23.92){\circle*{0.4}}
\put(15.50,8.00){\circle*{0.4}}
\put(17.58,8.83){\circle*{0.4}}
\put(19.42,10.17){\circle*{0.4}}
\end{picture}
\hfill\null
\caption{Indices on the splitted graph\label{raddisc}}
\end{figure}
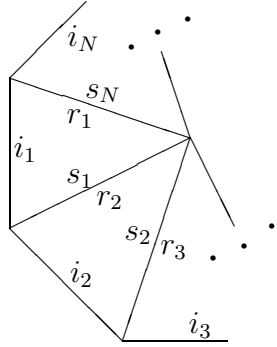

We consider the splitted graph, let $i_1$ be an arbitrary index. Let
$e_{i_1}\in R$, we consider all triangles which contain the vertex opposite
to the index $i_1$. We label the indices as in fig. \ref{raddisc}. The
partition function of this part of the graph, summed over all inner indices
$r_1,\ldots,r_N$ and $s_1,\ldots,s_N$, is given by
\BEAS
   Z_{i_1,\ldots,i_N}&=&\lambda_{r_1i_1s_1}q^{s_1r_2}\lambda_{r_2i_2s_2}
   \ldots\lambda_{r_Ni_Ns_N}q^{s_Nr_1}~=~
   \lambda_{r_1i_1}^{r_2}\lambda_{r_2i_2}^{r_3}\ldots
       \lambda_{r_Ni_N}^{r_1}\\
   &=& (R_{i_N}\cdot R_{i_{N_1}}\cdot\ldots\cdot R_{i_1} e_{r_1})^{r_1}~=~
       (R_{e_{i_1}e_{i_2}\ldots e_{i_N}}e_{r_1})^{r_1}\\
   &=& \tr R_{e_{i_1}e_{i_2}\ldots e_{i_N}}
\EEAS
where $R_a$ is the right multiplication in \alA\ considered as an
endomorphism: $R_a b=ba$, $R_i$ is short for $R_{e_i}$.
If $e_{i_1}\in R$, then is also $e_{i_1}e_{i_2}\ldots e_{i_N}\in R$ and the
trace vanishes, therefore all configurations with an index in $R$ give no
contribution to the partition function.

Now let $e_i\in L$, then $q^{ij}=0$ for all $e_j\not\in R$, but if
$e_j\in R$ then we can repeat the discussion above with the result
that also all configurations with an index in $L$ give no contribution
to the partition function.

There remains the discussion of the semisimple algebra \alB.
It is $\theta$-invariant and the (orthogonal) direct sum of all simple
ideals of \alA. One might expect that each simple ideal $I$ is itself
$\theta$-invariant, but this is not true in general.

$\theta$ is an automorphism of \alA, the image of a simple ideal $I_1$ is
also a simple ideal $I_2=\theta(I_1)$ which can be different from $I_1$.
We get a sequence $I_1,I_2,\ldots,I_k$ of disjoint isomorphic simple ideals
with $\theta(I_k)=I_1$; since $\theta^p=1$ the number $k$ must be a divisor
of $p$, $p=kl$. Not each ideal $I_i$ is $\theta$-invariant, but the direct
sum $I=I_1\oplus\ldots\oplus I_k$ is, and we can decompose it in
$I=I^{(0)}\oplus\ldots\oplus I^{(p-1)}$ with $I^{(j)}\subset \alA_j$ and
$\theta(I^{(j)})=\omega^j I^{(j)}$. The partition function decomposes into
several parts belonging to $\theta$-invariant semisimple ideals of \alA.

By the assumptions in theorem \ref{allgstruktur} $q|_{I^{(1)}}$ is non
degenerate. We will test this condition to gain information about $k$:
let $a,b\in I^{(1)}$, $a=a_1+\ldots+a_k$, $b=b_1+\ldots+b_k$, $a_j,b_j\in
I_j$. Since $\theta(a)=\omega a$ and $\theta(a_j)\in I_{j+1}$ we get
$\theta(a_j)=\omega a_{j+1}$, $\theta(a_k)=\omega a_1$ and therefore
\BEAS
   a_j~=~ \omega^{1-j}\theta^{j-1}(a_1)~~\Rightarrow~~a
      &=&\sum_j \omega^{1-j}\theta^{j-1}(a_1)\\
   b &=&\sum_j \omega^{1-j}
           \theta^{j-1}(b_1)
\EEAS
\BEAS
\Rightarrow q(a,b)&=&\sum_{i,j}\omega^{2-i-j} q(\theta^{i-1}(a_1),
      \theta^{j-1}(b_1))~=~
      \sum_i \omega^{2(1-i)} q(\theta^{i-1}(a_1),\theta^{i-1}(b_1))\\
     &=& \sum_i \omega^{2(1-i)} \omega^{2(i-1)} q(a_1,b_1)~=~
         k q(a_1,b_1)
\EEAS
where we have used $q(\theta(a),\theta(b))=\omega^2 q(a,b)$.
$\theta^k$ is an automorphisms of $I_1$, which is a simple complex algebra
isomorphic to a full complex matrix algebra. By the theorem of Noether-Skolem
(\cite{Pierce}) is $\theta^k$ an inner automorphism, i.e. there exists an
invertible element $s\in I_1$ with $\theta^k(a)=s^{-1}as$ for all $a\in I_1$.
Then
\BEAS
   q(\theta^k(a_1),\theta^k(b_1)&=& \omega^{2k}q(a_1,b_1)\\
   =~q(s^{-1}a_1s,s^{-1}b_1s) &=& q(a_1,b_1)~~~~\forall a_1,b_1\in I_1
\EEAS
where we have used the invariance and the symmetry of $q$.
Hence $\omega^{2k}=1$ which is only possible for $2k=p$ or $k=p$.
All other cases, e.g. $k=1$ for $p>2$ which corresponds to a
$\theta$-invariant simple ideal do not occur in the context of flip
invariant models.

There remains the discussion of this two cases:

$k=p$: This is the trivial one.
There are $p$ simple ideals isomorphic to a full complex matrix algebra
$\setC^{r\times r}$
Let $\{e_i\}$ be a basis of $I_1$, then is $\{\tilde{e}_i=e_i+
\omega^{-1}\theta(e_i)+\ldots+\omega^{1-p}\theta^{p-1}(e_i)\}$ a basis
of $\alA_1$. Denote by $(a)_1$ the $I_1$ component of $a$, then we get for
the weights
\BEA
    \Gamma_{i_1\ldots i_n} &=& q(\Gamma(\tilde{e}_{i_1},\ldots,
      \tilde{e}_{i_{p+1}}),\tilde{e}_{i_n})~=~
     q(\tilde{e}_{i_1}\ldots\tilde{e}_{i_{p+1}},\tilde{e}_{i_n})\nonumber\\
    &=&k q^{(1)}((\tilde{e}_{i_1}\ldots\tilde{e}_{i_{p+1}})_{1},
          (\tilde{e}_{i_n})_{1}) ~=~
     k q^{(1)}(e_{i_1}\ldots e_{i_{p+1}},e_{i_n}) \label{weightdecomp}
\EEA
This is exactly the weight one would get for a $n$-gon glued together out
of $n-2$ triangles with a topological weight on the triangles. Therefore
this case is called trivial.

For the calculation of the partition function it is convenient to
consider the dual graph in the double line representation \cite{BFN}.
We choose in $I_1$, which is isomorphic to a full complex matrix algebra,
the standard basis $\{E_{ij}\}$ of $r\times r$ matrices with
$(E_{ij})_{kl}=\delta_{ik}\delta_{jl}$. $q^{(1)}=q|_{I_1\times I_1}$
is an invariant metric on $I_1$, this is, up to a factor, the trace of
the matrices: $q^{(1)}(a,b)=\beta\tr(ab)$. We get for the weights of the
vertices of degree $n$
\BE
   \Gamma_{i_1j_1i_2j_2\ldots i_nj_n}=p\beta \delta_{j_1i_2}\delta_{j_2i_3}
   \ldots\delta_{j_ni_1}
\EE
and for the weights of the edges ($q(\tilde{a},\tilde{b})=pq^{(1)}(a,b)$)
\BE
   q^{i_1j_1i_2j_2}=(p\beta)^{-1}\delta_{j_1i_2}\delta_{j_2i_1}
\EE

\begin{figure}[hbt]
\null\hfill
\unitlength=3mm
\begin{picture}(13.42,15.92)(0,3)
\put(3.00,5.00){\line(5,2){5.00}}
\put(8.00,7.00){\line(0,1){5.00}}
\put(8.00,12.00){\line(-5,1){5.00}}
\put(8.33,6.08){\line(-5,-2){5.25}}
\put(8.33,6.08){\line(-1,-5){0.75}}
\put(9.25,6.17){\line(-1,-5){0.75}}
\put(9.25,6.17){\line(4,-1){3.25}}
\put(9.17,7.17){\line(5,-1){3.75}}
\put(9.17,7.17){\line(0,1){4.83}}
\put(9.17,12.00){\line(4,1){4.25}}
\put(9.17,13.00){\line(4,1){4.08}}
\put(9.17,13.00){\line(0,1){2.92}}
\put(8.00,15.92){\line(0,-1){2.92}}
\put(8.00,13.00){\line(-5,1){4.25}}
\put(5.50,12.42){\makebox(0,0)[rt]{$i_1$}}
\put(7.83,9.50){\makebox(0,0)[rc]{$i_1$}}
\put(5.58,6.17){\makebox(0,0)[rb]{$i_1$}}
\put(5.00,4.50){\makebox(0,0)[lt]{$i_2$}}
\put(7.67,3.42){\makebox(0,0)[rb]{$i_2$}}
\put(8.83,3.75){\makebox(0,0)[lt]{$i_3$}}
\put(11.17,5.50){\makebox(0,0)[rt]{$i_3$}}
\put(11.00,6.92){\makebox(0,0)[lb]{$i_4$}}
\put(9.33,9.58){\makebox(0,0)[lc]{$i_4$}}
\put(11.67,12.50){\makebox(0,0)[lt]{$i_4$}}
\put(11.92,13.95){\makebox(0,0)[rb]{$i_5$}}
\put(9.33,15.17){\makebox(0,0)[lc]{$i_5$}}
\put(7.83,15.17){\makebox(0,0)[rc]{$i_6$}}
\put(5.25,13.67){\makebox(0,0)[lb]{$i_6$}}
\end{picture}
\hfill\null
\caption{Double line representation with equal indices\label{doublegraph}}
\end{figure}
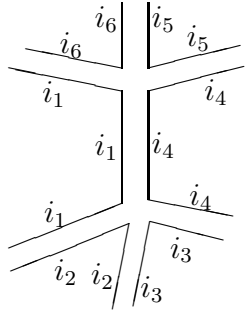

All indices on a closed line must have the same value
as indicated in fig. \ref{doublegraph}, each closed line
corresponds to a vertex of the original graph. The computation of the
partition function is therefore reduced to a counting of factors. We get
a factor $r$ for each vertex of the polygonization, a factor
$(p\beta)^{-1}$ for each edge and a factor $p\beta$ for each $n$-gon.
This results in
\BE
   Z~=~r^V(p\beta)^{-E}(p\beta)^P~=~(p\beta)^{\chi}(\frac{r}{p\beta})^{V}
\EE
where $V$ is the number of vertices of the polygonization, $E$ ist the
number of edges and $P$ the number of plaquettes, the $n$-gons.
We get the typical dependence of the partition function on the euler
characteristic $\chi$ of the manifold. If one adjusts the constant
$\beta$, such that $r=p\beta$, then the partition function will be
topological.

The other case $2k=r$ is non trivial and leads to totally new aspects.
Let $a=a_1+\ldots+a_k\in I^{(1)}$, i.e. $\theta(a)=\omega a$. Then
$a_i=\omega^{1-i}\theta^{i-1}(a_1)$, $\theta^k(a_1)=\omega^k a_1=-a_1$.
In this case is $\Theta=\theta^k$ an automorphism from $I_1$ to $I_1$
with $\Theta^2=1$. By the theorem of Noether-Skolem it is an inner
automorphism, there exists a $s\in I_1$ with $\Theta(a)=s^{-1}as$ for
all $a\in I_1$. Then $\Theta^2(a)=s^{-2}as^2=a$ i.e. $[a,s^2]=0$ for
all $a\in I_1$. With Schur's Lemma we conclude that $s^2=\lambda 1$, we
can set $\lambda=1$. Then we can choose a basis in $I_1$ such that
$ s={\rm diag}(1,\ldots,1,-1,\ldots,-1)$ with $M$ times $1$ and $N$ times
$-1$, $M+N=r$. $\Theta(a_1)=s^{-1}a_1 s=- a_1$ is fulfilled for all
matrices $a_1\in I_1$ which have the off diagonal block form
\BE
   a_1~=~\left(\begin{array}{cc} 0 & \Box \\ \Box & 0 \end{array}\right)
\EE
i.e. $(a_1)_{ij}=0$ for $i,j\leq M$ or for $i,j> M$. $I_1$ is then a
$\setZ_2$-graded algebra. A basis of $I^{(1)}$ is
given by $\{\tilde{E}_{ij}=E_{ij}+\omega^{-1}\theta(E_{ij})+\ldots+
\omega^{1-k}\theta^{k-1}(E_{ij})| i\leq M<j~or~j\leq M<i\}$. The weights are
given by
\BEA
   \Gamma_{i_1j_1i_2j_2\ldots i_nj_n}&=&k\beta \delta_{j_1i_2}\delta_{j_2i_3}
      \ldots\delta_{j_ni_1}\\
   q^{i_1j_1i_2j_2} &=& (k\beta)^{-1}\delta_{j_1i_2}\delta_{j_2i_1}
\EEA
where the pairs $i_1j_1,i_2j_2,\ldots,i_nj_n$ fulfil alternating the
relations $i\leq M<j$ and $j\leq M<i$.

\begin{figure}[hbt]
\null\hfill
\unitlength=3mm
\begin{picture}(13.42,15.92)(0,3)
\put(3.00,5.00){\line(5,2){5.00}}
\put(8.00,7.00){\line(0,1){5.00}}
\put(8.00,12.00){\line(-5,1){5.00}} \thicklines
\put(8.33,6.08){\line(-5,-2){5.25}}
\put(8.33,6.08){\line(-1,-5){0.75}} \thinlines
\put(9.25,6.17){\line(-1,-5){0.75}}
\put(9.25,6.17){\line(4,-1){3.25}} \thicklines
\put(9.17,7.17){\line(5,-1){3.75}}
\put(9.17,7.17){\line(0,1){4.83}}
\put(9.17,12.00){\line(4,1){4.25}} \thinlines
\put(9.17,13.00){\line(4,1){4.08}}
\put(9.17,13.00){\line(0,1){2.92}}  \thicklines
\put(8.00,15.92){\line(0,-1){2.92}}
\put(8.00,13.00){\line(-5,1){4.25}}
\put(5.50,12.42){\makebox(0,0)[rt]{$i_1$}}
\put(7.83,9.50){\makebox(0,0)[rc]{$i_1$}}
\put(5.58,6.17){\makebox(0,0)[rb]{$i_1$}}
\put(5.00,4.50){\makebox(0,0)[lt]{$i_2$}}
\put(7.67,3.42){\makebox(0,0)[rb]{$i_2$}}
\put(8.83,3.75){\makebox(0,0)[lt]{$i_3$}}
\put(11.17,5.50){\makebox(0,0)[rt]{$i_3$}}
\put(11.00,6.92){\makebox(0,0)[lb]{$i_4$}}
\put(9.33,9.58){\makebox(0,0)[lc]{$i_4$}}
\put(11.67,12.50){\makebox(0,0)[lt]{$i_4$}}
\put(11.92,13.95){\makebox(0,0)[rb]{$i_5$}}
\put(9.33,15.17){\makebox(0,0)[lc]{$i_5$}}
\put(7.83,15.17){\makebox(0,0)[rc]{$i_6$}}
\put(5.25,13.67){\makebox(0,0)[lb]{$i_6$}}
\end{picture}
\hfill\null
\caption{Double line representation of a chequered graph\label{doublecheque}}
\end{figure}
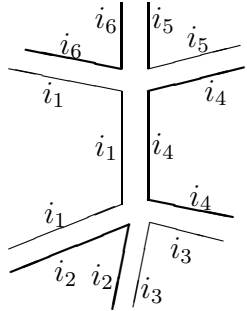

In the double line representation of the dual graph each double line carries
both types of indices, the indices of a line must have the same value.
See fig. \ref{doublecheque} where different linetypes denote different
ranges of indices.

For an arbitrary graph it is not possible to distribute the indices in
this manner. In case it is the graph is called chequered \cite{Morris,BFN},
i.e. the faces of the dual graph can be coloured alternating black and
white such that nowhere are two black or two white faces are neighboured.
If the graph is not chequered the partition function vanishes, otherwise
we get
\BE
   Z=(M^{V_1}N^{V_2}+M^{V_2}N^{V_1})(k\beta)^{P-E}
\EE
where $V_1$ and $V_2$ are the numbers of vertices whose dual plaquettes
carry the same type of index, these are flip invariants of the model.

This models can distinguish smaller classes of graphs, the flip move is
therefore not transitive. See \cite{BFN} for a discussion of the
consequences of this fact.

\end{document}